# Gastrointestinal Mucosal Problems Classification with Deep Learning

Mohammadhasan Goharian[1], Vahid Goharian[2], Hamidreza Bolhasani[3*]

*Abstract*— Gastrointestinal mucosal changes can cause cancers after some years and early diagnosing them can be very useful to prevent cancers and early treatment. In this article, 8 classes of mucosal changes and anatomical landmarks including Polyps, Ulcerative Colitis, Esophagitis, Normal Z-Line, Normal Pylorus, Normal Cecum, Dyed Lifted Polyps, and Dyed Lifted Margin were predicted by deep learning. We used neural networks in this article. It is a black box artificial intelligence algorithm that works like a human neural system. In this article, Transfer Learning (TL) based on the Convolutional Neural Networks (CNNs), which is one of the well-known types of neural networks in image processing is used. We compared some famous CNN architecture including VGG, Inception, Xception, and ResNet. Our best model got 93% accuracy in test images. At last, we used our model in some real endoscopy and colonoscopy movies to classify problems.

*Keywords*- Deep Learning, Convolutional Neural Networks, Medical Image Classification, Endoscopy, Colonoscopy, Gastroenterology, Polyps, Colorectal Cancer, Transfer Learning

## I. INTRODUCTION

Gastrointestinal problems are one of the most common diseases in the world. mucosal changes in the digestive system are very helpful in diagnosing. Endoscopy is used to diagnose these mucosal changes. Some of these mucosal changes like esophagitis [1], inflammatory bowel diseases [2] and polyps [3] can cause gastrointestinal cancers. Colorectal Cancer (CRC) is the third most common cancer and the second leading cause of cancer related deaths in the world with an estimated number of 1.8 million new cases and about 881,000 deaths worldwide in 2018 [4]. Early diagnosis of mucosal changes can help for cheaper and more effectively treatment, also decrease death probability. Detecting some dyed mucosal place can also be very helpful, because medical doctors and specialists use them as landmarks for surgery and after surgery, they should notice dyed margins as a measure of surgical success. Endoscopy has two ways, upper endoscopy and lower endoscopy (colonoscopy) and it's the best way to accurately diagnose mucosal changes. Human factors, like visual errors can make some error in diagnosis also humans need more time for watch and find problems. Deep learning can help humans to do endoscopy more accurate and more quickly. New studies [12-16] have shown neural networks were able to show high accuracy in recognizing endoscopic images. What has not been shown in these studies is that it can detect several mucosal changes and different problems with a single neural network with high accuracy.

Neural Network is a subset of artificial intelligence that have been made according to the structure of the human nervous system and they are more powerful than classical machine learning in solving problems, especially in big data and data with more dimensions such as signals and images. One of the abilities of the neural networks is that they can recognize the effective features inside the photos like edges, colors, and different margins in the unusual mucus compared to the normal mucus and make their own diagnosis about it.

Compared to the classical methods of machine learning, there is no need to extract features manually, as a result, the network is more fitted base on the dataset, and according to different shows of mucosal problems in each patient, even if the type of disease is the same, the hidden relationships in the data is found by network and the accuracy of our model increases. When the proposed network moves from first to the last layers, the features evolve more, for example, they recognize an edge in first layer and compound some of them for second layers and finally they can fully recognize a polyp in the last layers.

[1] School of Medicine, Islamic Azad University Medical Sciences Najaf Abad, Isfahan, Iran.
[2] School of Medicine, Isfahan University of Medical Sciences, Isfahan, Iran.
[3] Department of Computer Engineering, Science and Research Branch, Islamic Azad University, Tehran, Iran.
* Corresponding Author: hamidreza.bolhasani@srbiau.ac.ir

We owe the growth of accuracy of neural networks in photos to Convolutional Neural Networks (CNNs), which use weights or multi-dimensional filters to be able to establish the relationship between the rows and columns of the photos, which can be very useful in problems with related pixels such as esophagitis and polyps. Due to the number and type of layers of neural networks, there are different architectures in neural networks that we have used in this article to diagnose gastrointestinal tract problems and we compare their accuracy to choose the best architecture for this work.

Also, if a network has this ability to detect various diseases and problems that a person may suffer from in real time, it can appear faster in screening tests compared to some networks. This can make a person, aware of his health condition and refer him to the doctor for the next tests and treatment. For this purpose, in this article it was tried to predict many different problems and situations in the Kvasir dataset [5] in only one network, and as a result, unlike some similar articles, we did not use some networks to distinguish each category from other categories.

Esophagitis, Polyp, Ulcerative Colitis, Normal Z-Line, Normal Cecum, Normal Pylorus, Dyed Lifted Polyp, Dyed Lifted Margin are different categories of Kvasir dataset and in this research, they were predicted by different neural networks architects and we compared to find best neural network for this work.

The organization of this paper is presented in this order: Section 2 covers a compact review on the related works. In Section 3, background and motivations of this research are offered. Then materials and methods including data-preprocessing, and the proposed model is discussed. Study results in the form of diagrams and tables are described in Section 5. And conclusion is included in section 6.

## II. RELATED WORKS

Compared to some similar works like [5] and [6], at first, we used data augmentation that means we flipped and rotated our images. It can make predictions, more difficult but it is more practical in the real works and in addition we obtained about 4% more accuracy with data augmentation compared to [6] article. Secondly, unlike those two articles, we split our data to three parts, train, validation, and test, so we tested our models in test, not validation. It can reduce accuracy, but again we got around same accuracy and it is more practical in the real works. Thirdly, in this article, recall and precision for each class and each network was calculated, so based on the risk level of each class and the type of endoscopy (upper or lower), we can choose the best network in real works. This model was also tested with some movies to make it a new product in the near future. Unlike [7], we split our test dataset from the beginning and it is like new data for our network so it can work better in the practical work.

## III. BACKGROUND AND MOTIVATIONS

In this particle, Kvasir dataset was used, including 8,000 endoscopy and colonoscopy images. It can be very helpful to diagnose patients with higher probability of risk of cancer and treat their disease to prevent cancer. Kvasir has 8 categories for classification. Three of them are considered as normal: Z-Line, Normal Pylorus and Normal Cecum. Figure 1, shows the appearance and anatomical landmarks of this category that considered as normal populations.

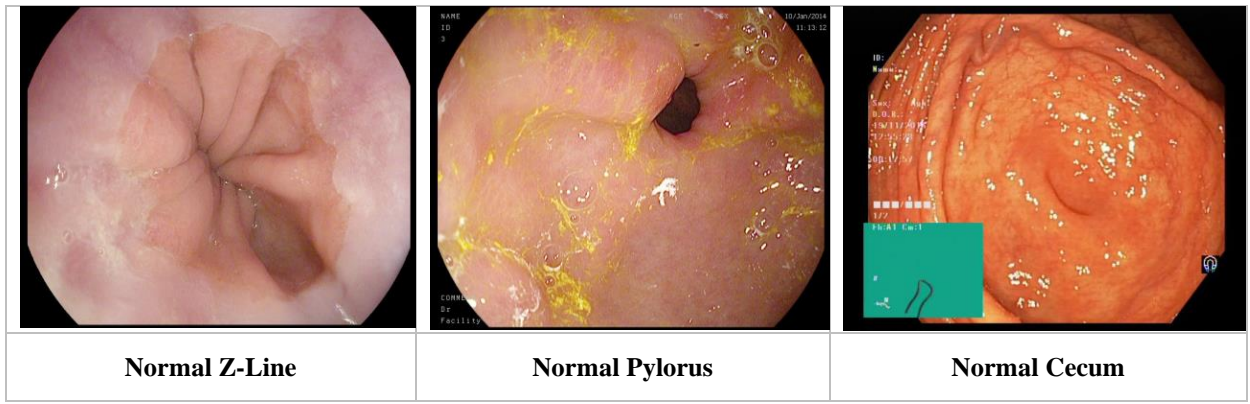

Fig. 1. Normal categories in Kvasir dataset [5] with some image samples

Three of them are Esophagitis, Polyps and Ulcerative Colitis (Fig. 2). These are clinical important findings. We count them as pathological findings and damages. In the patient's early detection of these problems, they can be very useful on medical screening tests.

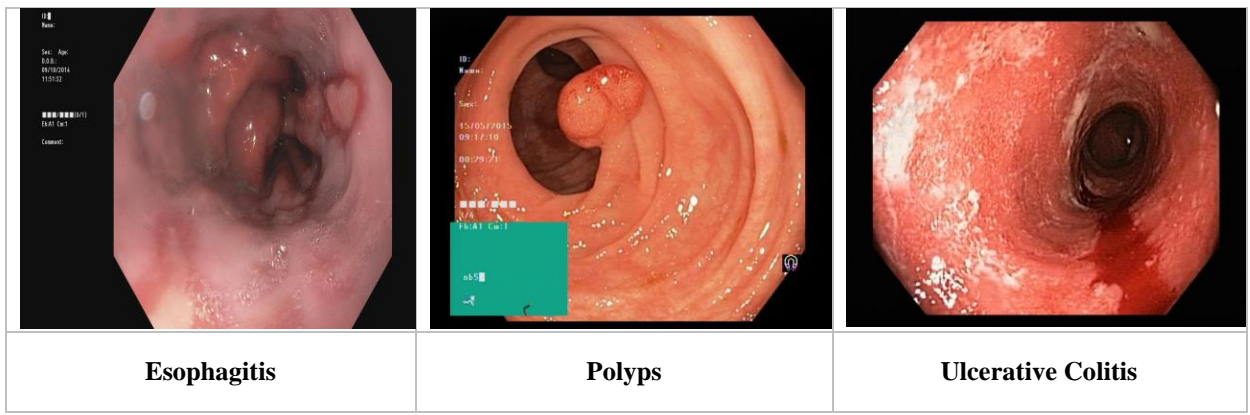

Fig. 2. Image samples of Esophagitis, Polyps, and Ulcerative Colitis

Two last categories are dyed lifted polyp and dyed resection margins (Fig. 3). These two categories are used for polyp removal and can help surgeons for surgery.

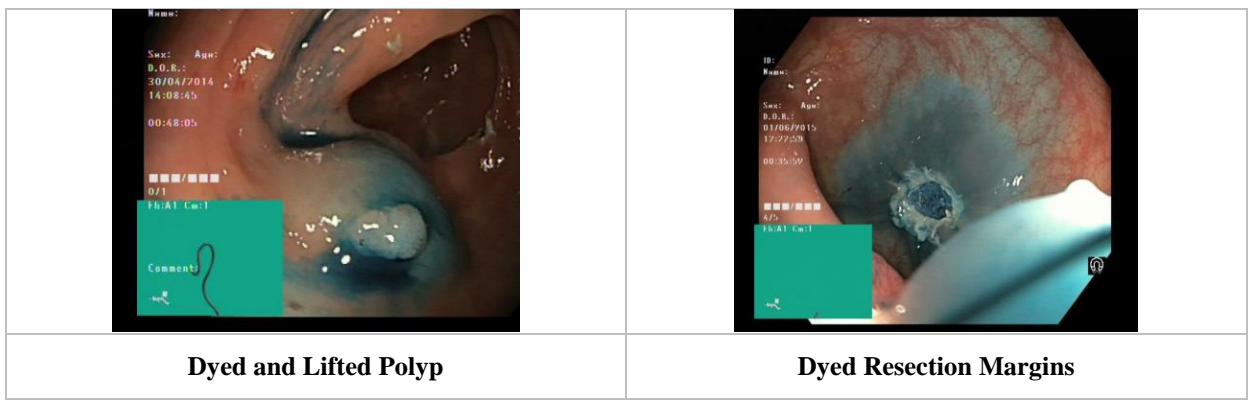

Fig. 3. Image samples of Dyed Lifted Polyp and Dyed Resection Margins

A lot of gastrointestinal patients do not have obvious clinical signs and a fast and reliable screening test, but endoscopy can be very useful for early diagnosis. Kvasir dataset link: https://datasets.simula.no/kvasir/

**Convolutional Neural Networks (CNNs)**

CNN is considered as one of the well-known deep neural networks. It is commonly used for computer vision tasks. In CNNs, convolutional and max-pooling layers are used. Convolutional layers using 2D filters with 3*3 dimensions, 5*5 dimensions and above but for colored images, 3D filters are used. In each place in filters, a weight should be trained for each specific network. Convolutional layers can extract features of the image. Max-pooling layers are used for dimension reduction and increase train and predict speed. After them, a fully connected layer (FCN) is used for predict from extracted features from CNN. In the end, a Softmax layer can be useful to predict probability of belonging to each class.

**- VGG19**

The architecture [8], created by visual geometry group from Oxford, this architecture receives 224*224 RGB photo input and RGB includes 3 red green blue channels so it actually takes 224*224*3 dimensions, in the first and second layers, it uses 64 filters with 3*3 size. Then it reduces the size with max pool 2*2 for faster processing, the fourth and fifth layers include 128 3*3 filters and the sixth layer includes max pool 2*2, the next four layers are 256 3*3 filters in each layer, the eleventh layer is a 2*2 max pool layer, which reduces the dimensions of the features, then four layers with 512 filters with 3*3 size are used, one max pooling, and again 4 layers with 512 filters and another max pool, at the end it has 3 fully connected network layer for detection based on the features extracted by the filters and the Softmax layer, which gives us the probability of belonging to each category, so VGG has total of 16 convolutional layers, 5 Maxpool layers, 3 fully connected layers, and 1 Softmax layer.

**- Inception**

It is worth noting that in CNNs, filters with higher dimensions such as 7*7 and 5*5 have more weights. As a result, they work more powerfully, but because of the many parameters, and the number of weights they should learn, they have slow speed in learning, instead in 1*1 and 3*3 filters have higher speed. In the Inception architecture [9], there are blocks (the name of which is taken from the Inception movie) that uses 1*1, 3x3 and 5x5 filters that can extract features faster (by lowering the parameter), and also it is more powerful. In each inception block, there are 6 convolutional layers and one Maxpool layer. The whole network includes 3 convolutional layers, 4 Maxpool layers, and 3 inception blocks, and at the end, fully connected layers and Softmax is used to detect the class from the extracted features.

**- Xception**

The architecture was created by Francois Chollet from Google. Xception [10] means extreme inception. Xception takes the principle of inception but does it inverse. In the inception block, the feature map is first compressed, and then the filters are applied with high dimensions. Here, the filters are applied first, and then, the compressed input with a 1x1 filter is added to the output.

**- ResNet**

ResNet or residual network architecture was built by Kaming He, and Xiangyu Zhang in 2015, won the ILSVRC[4] 2015 competition. It was inspired by VGG19 architecture. In the neural network, for a large number of parameters, there is a possibility of becoming zero due to the optimization with gradient descent method, and as a result, this problem can cause the loss of information in higher layers. To solve this problem, ResNet gives a weight to inputs of each layer and adds that to output to make sure each layer can work at least as good as the last layer, that is why it is called ResNet. It contains 34 convolution layers, Maxpool, fully connected and Softmax, which Softmax shows probability of belonging to each of 8 Classes (Polyp, Ulcerative Colitis, Normal Pylorus, Normal Z-Line, Esophagitis, Normal Cecum, Dyed Lifted Polyp and Dyed Resection Margin).

## IV. MATERIALS AND METHODS

In this this research, the performance of VGG, Inception, Xception, ResNet was tested for training the network on the Kvasir dataset. Before training the network, some pre-processing on the data was applied to remove noises and better performance. Data was divided into two train and test groups to check different networks on it to compare the performance of different architectures. Also, we applied data augmentation to the data for better performance of the network in different situations. Finally, we trained the network. Kvasir dataset is a comprehensive dataset with 8000 images from the size of 720*576 to 1920*1072 pixels. As mentioned, these 8000 images are divided into 8 classes: Ulcerative Colitis, Normal Cecum, Polyps, Dyed Lifted Polyp, Dyed Lifted Margin, Normal Pylorus, and Normal Z-Line.

**Data Pre-Processing**

In the beginning, we resize the images into 128*128 pixels, because of the same size for neural network inputs and if we want to use it in practical work, we can analyze the images with lower resolution for some older devices. In this work, intra-area method was used for resizing, which is doing resize with pixel area relation. Then, all of the pixels were divided to 255, now our pixels are between 0 and 1. It helps for network stability because without this step, we have higher difference between our weights and pixels and a little difference in our input can make a big change in our result.

**Test and Train Split**

In this work, features and classes divided into two parts, train 90% and test 10%. Then we divided train into two parts. Again, validation 20% and train 80%, validation is for testing network with different weights and choose the least loss. If we use test dataset directly, it can cause overfit and insensitivity in test dataset.

**Data Augmentation**

To get better results in practical work and simulating the real environment, we made input data, more difficult for our network. We used rotation, shear, zoom, width-shift, horizontal-shift and horizontal flip. Now if our network gets inputs with different angel ulcers or nearer endoscopy camera or different place of landmarks, it can predict better and we have the better diagnosis.

We have 16 images in each of our batches. It helps for easier training and in comparison, with stochastic it helps for higher speed and lower noise in our optimization for weights.

---

[4] ILSVRC: ImageNet Large Scale Visual Recognition Challenge

## Model

In our study, the first layer is batch normalization. It makes the distribution of data the same in each batch. Then, we use the body of different pre-trained networks, ResNet152v2, VGG19, InceptionV3 and Xception. After that, 128 CNN filters with 1*1 shape were used in the outputs of that network. The flattened output of the CNN is our extracted features of images, that can be the highest level of images we extracted and two fully connected layers with 512 and 8 neuron was used to get the final output. At the end, we used Softmax in order of helping us to predict the probability of belonging to each class (Fig. 4).

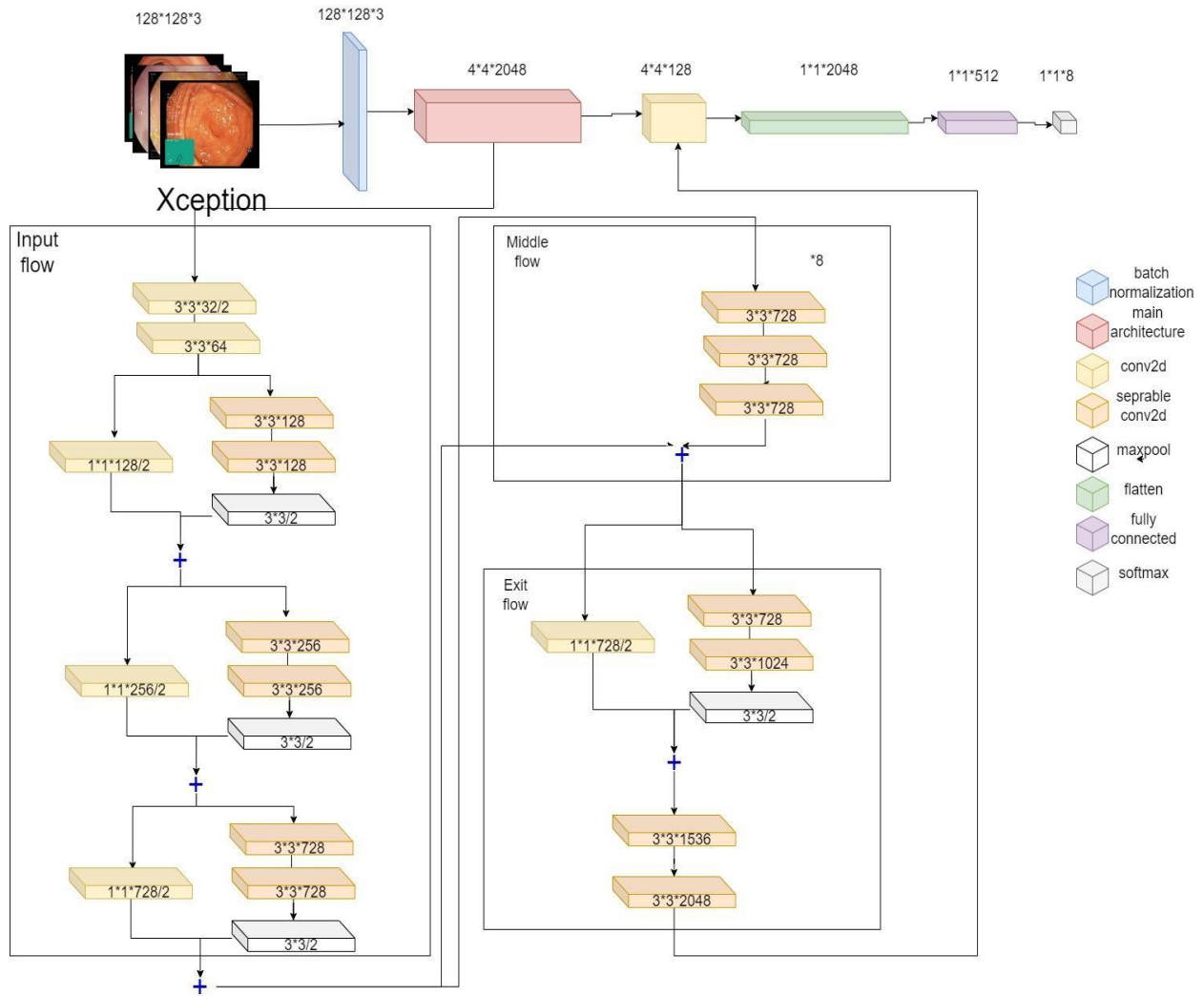

Fig. 4. The proposed model architecture

## Transfer Learning

For faster and better training, these networks are pre trained in ImageNet dataset and these weights as initial weights are better than random weights. These weights are fine-tuned for Kvasir dataset, that means weights updated and learned Kvasir classes.

**Loss and Optimization**

Defining loss help us to measure the model error, and optimization tries to decrease the loss to minimum value. In this study we used cross entropy loss (Equation 1):

$$Loss = - \sum_{i-1}^{output\ size} y_i . \log \hat{y}_i \quad (1)$$

And we used the mean of this loss for each batch for optimization.

In our model Adam optimization was used. It is an update of RMSprop and use it with momentum it can adapt with each gradient changing the learning rate for each one automatically. It helps optimization to prevent optimization from some noises, in order of updating parameters to reach to the best parameters with less epochs (Equations 2-5).

$$m_t = \beta_1 m_{t-1} + (1 - \beta_1) g_t \quad (2)$$

$$v_t = \beta_2 v_{t-1} + (1 - \beta_2) g_t^2 \quad (3)$$

$$\hat{v}_t = max(\hat{v}_{t-1}, v_t) \quad (4)$$

$$\theta_{t+1} = \theta_t - \frac{\eta}{\sqrt{\hat{v}_t + \epsilon}} m_t \quad (5)$$

M = Momentum
Beta1 and Beta2 = Decay Rate
$g_t$ = Gradient

## V. STUDY RESULTS

In this study, for evaluating the proposed model, F1-score, accuracy, precision, and recall were used. To compute them, confusion matrix is necessary that has four outcomes. True positive which shows the labels belong to positive class and model correctly predict them. True negative shows the labels belong to negative class and model correctly predict them. False positive represents labels belong to negative class and model incorrectly predict them as positive class and false-negative represents the labels belong to positive class and model incorrectly predict them as negative class (Equations 6-9).

$$Accuracy = \frac{TP + TN}{TP + TN + FP + FN} \quad (6)$$

$$Precision = \frac{TP}{TP + FP} \quad (7)$$

$$Recall = \frac{TP}{TP + FN} \quad (8)$$

$$F1-score = \frac{2 \times Precision \times Recall}{Precision + Recall} \qquad (9)$$

F1-score precision and recall were computed for each class separately (Table 2-5). Recall is more important for some diseases in medicine, since if diagnose them as a possibility with some other examination like pathology, we can become sure about problem. Therefore, in some classes like polyp, recall is more important than precision for us. Also, we can choose different model for upper and lower gastrointestinal (GI) base on separated outcomes but accuracy was computed for all of the classes in the test dataset together. Best model we trained, was Xception with 93% accuracy, second model was VGG19 with 92% accuracy, third model was ResNet, with 91% accuracy, and the last one was Inception with 90% accuracy (Table 1).

* For the implementation of this work, Google Colab with free GPU was used. The benefit of this approach is that everyone can test it with going to this site for free.

Table. 1. Accuracy of evaluated models

|  | **Inception** | **ResNet** | **VGG19** | **Xception** |
|---|---|---|---|---|
| **Accuracy** | 90% | 91% | 92% | 93% |

Table. 2. Performance results of Xception model

| **Xception** | **Precision** | **Recall** | **F1-Score** | **Support** |
|---|---|---|---|---|
| Normal Pylorus | 93% | 95% | 94% | 86 |
| Normal Z-Line | 95% | 93% | 94% | 89 |
| Dyed Resection Margin | 96% | 99% | 98% | 99 |
| Ulcerative Colitis | 76% | 92% | 83% | 87 |
| Dyed Lifted Polyp | 94% | 95% | 94% | 115 |
| Normal Cecum | 93% | 75% | 83% | 110 |
| Esophagitis | 96% | 94% | 95% | 117 |
| Polyps | 97% | 99% | 98% | 97 |

Table. 3. Performance results of VGG19 model

| VGG19 | Precision | Recall | F1-Score | Support |
|---|---|---|---|---|
| Normal Pylorus | 91% | 94% | 93% | 86 |
| Normal Z-Line | 95% | 89% | 92% | 89 |
| Dyed Resection Margin | 99% | 99% | 99% | 99 |
| Ulcerative Colitis | 76% | 93% | 84% | 87 |
| Dyed Lifted Polyp | 90% | 97% | 93% | 115 |
| Normal Cecum | 94% | 75% | 83% | 110 |
| Esophagitis | 94% | 91% | 93% | 117 |
| Polyps | 96% | 96% | 96% | 97 |

Table. 4. Performance results of ResNet152V2 model

| ResNet152V2 | Precision | Recall | F1-Score | Support |
|---|---|---|---|---|
| Normal Pylorus | 92% | 97% | 94% | 86 |
| Normal Z-Line | 84% | 88% | 90% | 89 |
| Dyed Resection Margin | 96% | 100% | 98% | 99 |
| Ulcerative Colitis | 78% | 79% | 79% | 87 |
| Dyed Lifted Polyp | 99% | 83% | 91% | 115 |
| Normal Cecum | 85% | 83% | 84% | 110 |
| Esophagitis | 93% | 95% | 94% | 117 |
| Polyps | 99% | 93% | 96% | 97 |

Table. 5. Performance results of Inception model

| Inception | Precision | Recall | F1-Score | Support |
|---|---|---|---|---|
| Normal Pylorus | 91% | 97% | 94% | 86 |
| Normal Z-Line | 95% | 89% | 92% | 89 |
| Dyed Resection Margin | 98% | 99% | 98% | 99 |
| Ulcerative Colitis | 75% | 83% | 79% | 87 |
| Dyed Lifted Polyp | 91% | 92% | 92% | 115 |
| Normal Cecum | 85% | 77% | 81% | 110 |
| Esophagitis | 93% | 85% | 89% | 117 |
| Polyps | 91% | 99% | 95% | 97 |

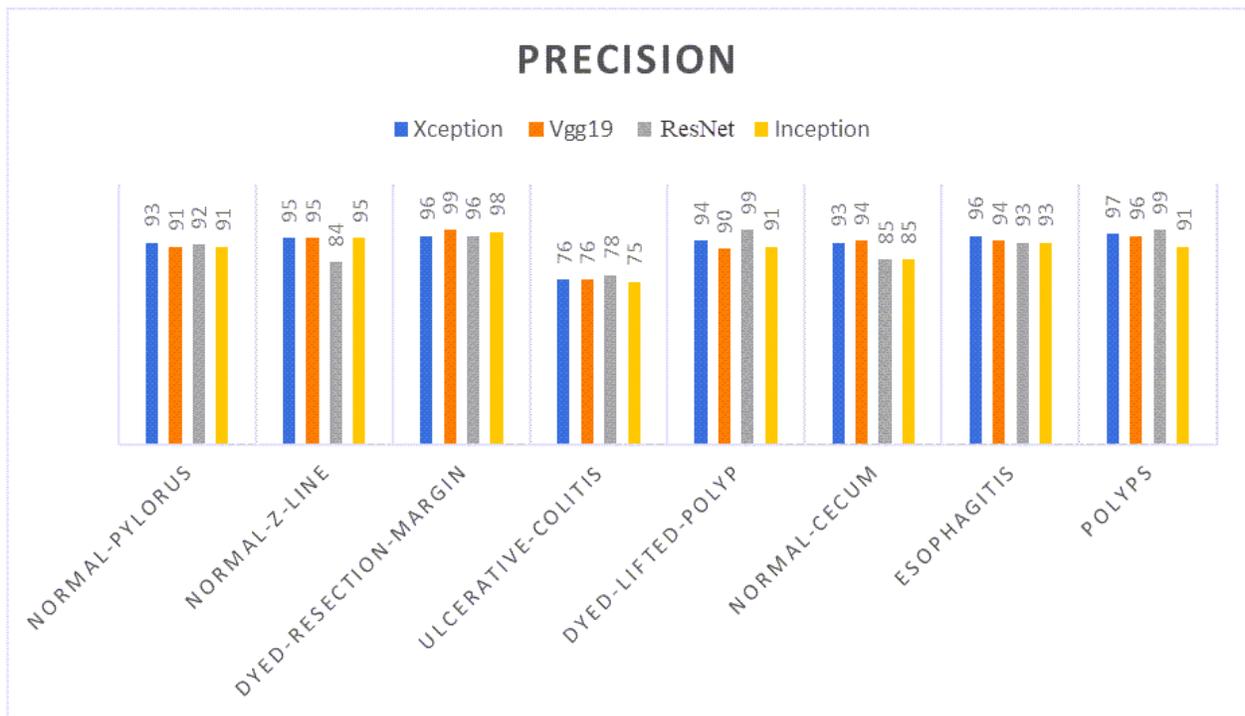

Fig. 5. Precision Results for Xception, VGG19, ResNet, and Inception

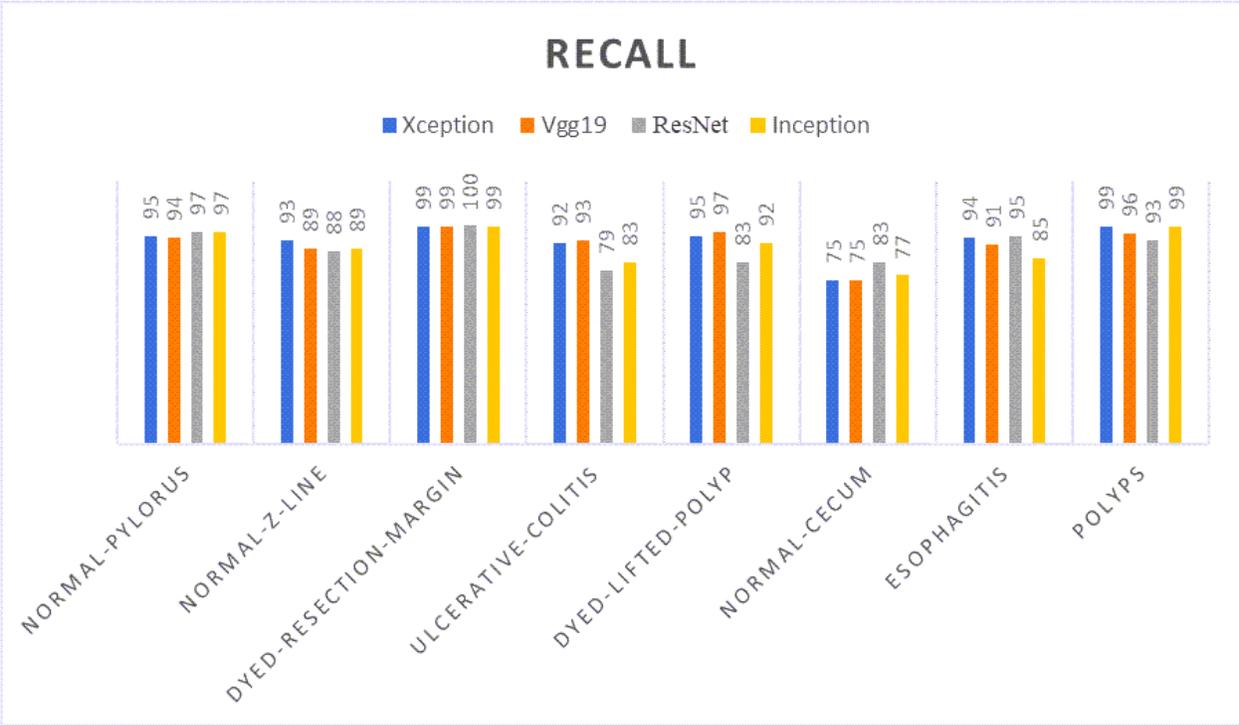

Fig. 6. Recall Results for Xception, VGG19, ResNet, and Inception

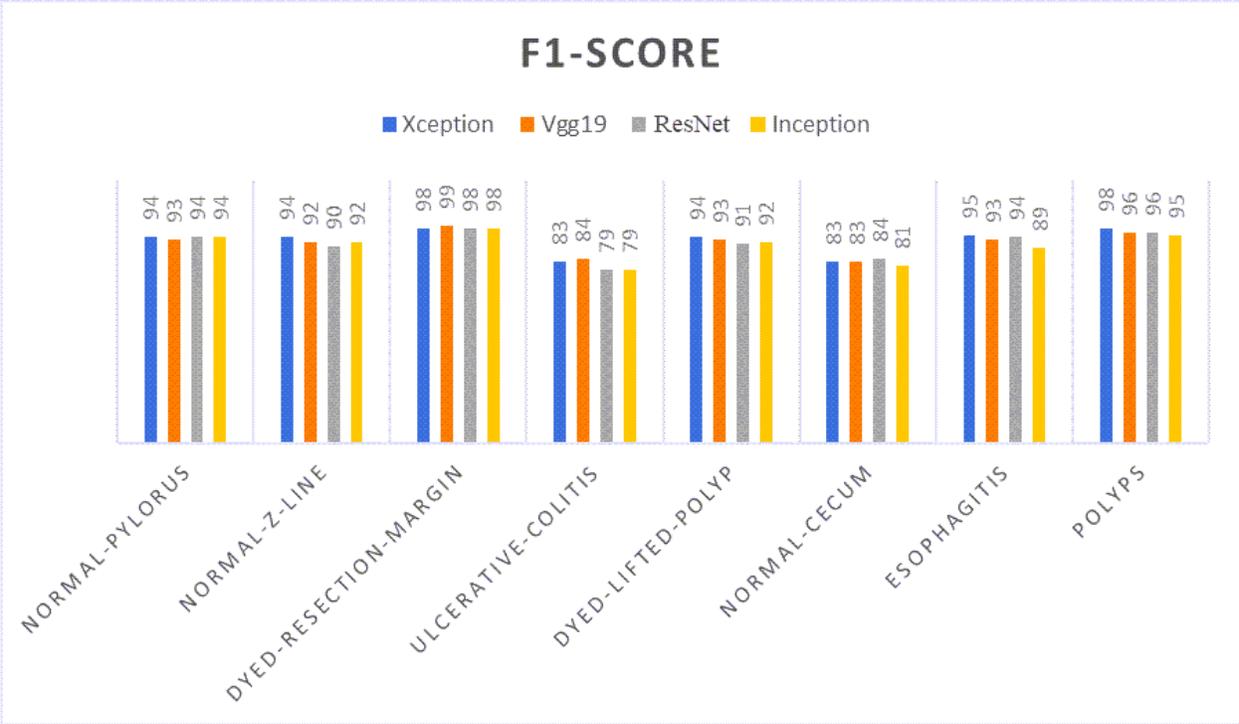

Fig. 7. F1-Score Results for Xception, VGG19, ResNet, and Inception

## VI. CONCLUSION

Early diagnosis in medicine can help so much to prevent more problems. If a polyp was not diagnosed, it can become cancer or ulcerative colitis, which can clearly increase the risk of cancer if it wasn't treated. Endoscopy is one of the most useful tests for detect GI tract problems. Medical specialists can become tired and due to human mistakes, deep learning can be very useful to detect problems. In this study, Kvasir dataset was used in 8 classes. We used Inception, Xception, VGG19 ResNet with pretrained weights and optimized them with Adam optimization to minimize the loss. According to the study results, the best model sorted as Xception, VGG19, ResNet, and Inception. Then we used some endoscopy and colonoscopy videos from different sources and cut them into frames, used our model on them. The results demonstrated that it can connect to an endoscopy device to help doctors and patients for a faster and more accurate diganosis.

## VII. ABBREVIATION

Table. 6. Abbreviations

| Abbreviation | Definition | Abbreviation | Definition |
| --- | --- | --- | --- |
| **CRC** | Colorectal Cancer | **TL** | Transfer Learning |
| **AI** | Artificial Intelligence | **CNN** | Convolutional Neural Network |
| **ML** | Machine Learning | **FCN** | Fully Connected Layer |
| **DL** | Deep Learning | **GI** | Gastrointestinal |
| **RMSprop** | Root Mean Squared Propagation | | |

## VIII. DECLARATIONS

AVAILABILITY OF DATA AND MATERIALS
Available.

FUNDING
Not Applicable.

ACKNOWLEDGEMENTS
Not Applicable.

CONFLICT OF INTEREST
o   All authors have participated in (a) conception and design, or analysis and interpretation of the data; (b) drafting the article or revising it critically for important intellectual content; and (c) approval of the final version.

- o This manuscript has not been submitted to, nor is under review at, another journal or other publishing venue.

- o The authors have no affiliation with any organization with a direct or indirect financial interest in the subject matter discussed in the manuscript

- o The following authors have affiliations with organizations with direct or indirect financial interest in the subject matter discussed in the manuscript:

## IX. REFERENCES

[1] Kuylenstierna, Richard, and Eva Munck-Wikland. "Esophagitis and cancer of the esophagus." Cancer 56.4 (1985): 837-839.
[2] Rogler, Gerhard. "Chronic ulcerative colitis and colorectal cancer." Cancer letters 345.2 (2014): 235-241.
[3] Noffsinger, Amy E. "Serrated polyps and colorectal cancer: new pathway to malignancy." Annual Review of Pathology: Mechanisms of Disease 4 (2009): 343-364.
[4] Baidoun, Firas, et al. "Colorectal cancer epidemiology: recent trends and impact on outcomes." Current drug targets 22.9 (2021): 998-1009.
[5] Pogorelov, Konstantin, et al. "Kvasir: A multi-class image dataset for computer aided gastrointestinal disease detection." Proceedings of the 8th ACM on Multimedia Systems Conference. 2017.
[6] Mukhtorov, Doniyorjon, et al. "Endoscopic image classification based on explainable deep learning." Sensors 23.6 (2023): 3176.
[7] Dheir, Ibtesam M., and Samy S. Abu-Naser. "Classification of anomalies in gastrointestinal tract using deep learning." (2022).
[8] Simonyan, Karen, and Andrew Zisserman. "Very deep convolutional networks for large-scale image recognition." arXiv preprint arXiv:1409.1556 (2014).
[9] Szegedy, Christian, et al. "Rethinking the inception architecture for computer vision." Proceedings of the IEEE conference on computer vision and pattern recognition. 2016.
[10] Chollet, François. "Xception: Deep learning with depthwise separable convolutions." Proceedings of the IEEE conference on computer vision and pattern recognition. 2017.
[11] He, Kaiming, et al. "Deep residual learning for image recognition." Proceedings of the IEEE conference on computer vision and pattern recognition. 2016.
[12] Jha, Debesh, et al. "Real-time polyp detection, localization and segmentation in colonoscopy using deep learning." Ieee Access 9 (2021): 40496-40510.
[13] Nogueira-Rodríguez, Alba, et al. "Real-time polyp detection model using convolutional neural networks." Neural Computing and Applications 34.13 (2022): 10375-10396.
[14] Pacal, Ishak, et al. "An efficient real-time colonic polyp detection with YOLO algorithms trained by using negative samples and large datasets." Computers in biology and medicine 141 (2022): 105031.
[15] Pacal, Ishak, and Dervis Karaboga. "A robust real-time deep learning based automatic polyp detection system." Computers in Biology and Medicine 134 (2021): 104519.
[16] Klare, Peter, et al. "Automated polyp detection in the colorectum: a prospective study (with videos)." Gastrointestinal endoscopy 89.3 (2019): 576-582.

**Authors Biography**

| | |
|---|---|
| **Mohammadhasan Goharian, Medical Student**<br><br>4th year Medical Student at School of Medicine, Islamic Azad University Medical Sciences Najaf Abad<br><br>Working in Deep Learning and Machine Learning with Python specially in Signal and Image Processing for 5 years<br><br>Interests: Using AI, especially Deep Learning in Health for Prediction, Diagnosis, and Treatment in different diseases | 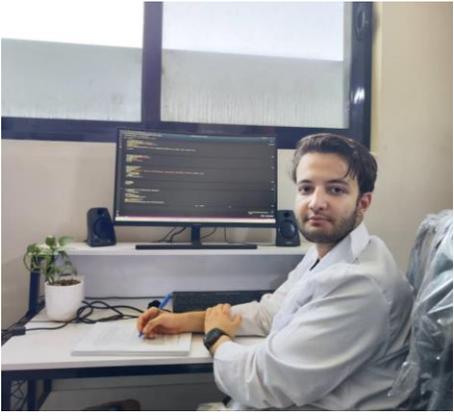 |
| **Vahid Gohrian, MD**<br><br>General Thoracic Surgery, MD<br><br>School of Medicine, Isfahan University of Medical Sciences, 2008-2010.<br><br>General Surgery / School of Medicine, Isfahan University of Medical Sciences, 2003-2007<br><br>General Practitioner / School of Medicine, Isfahan University of Medical Sciences, 1992-1999 | 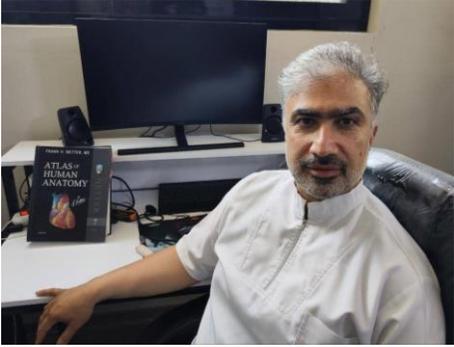 |
| **Hamidreza Bolhasani, PhD**<br><br>AI/ML Researcher / Visiting Professor<br><br>Founder and Chief Data Scientist at DataBiox<br><br>Ph.D. Computer Engineering from Science and Research Branch, Islamic Azad University, Tehran, Iran. 2018-2023.<br><br>Fields of Interest: Machine Learning, Deep Learning, Neural Networks, Computer Architecture, Bioinformatics | 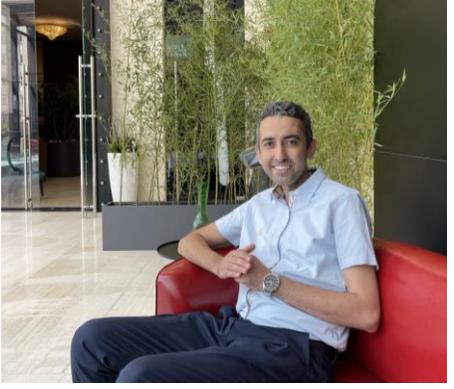 |